\documentclass[preprint,11pt]{elsarticle}

\usepackage[final]{changes}
\usepackage{setspace}
\usepackage{multirow,booktabs}
\usepackage{graphicx}
\usepackage[cmex10]{amsmath}
\usepackage{amsmath}
\usepackage{amssymb}
\usepackage{amsthm}

\usepackage{color}
\usepackage{numcompress}
\newcommand{\FigSize}{0.8\textwidth}
\newcommand{\FigSizeWide}{0.98\textwidth}

\journal{Applied Energy}

\begin{document}

\begin{frontmatter}

\title{Digital Twins based Day-ahead Integrated Energy System Scheduling under Load and Renewable Energy Uncertainties}

\author[label1]{Minglei You}
\author[label2]{Qian Wang}
 \author{Hongjian~Sun\corref{cor1}\fnref{label1}}
 \ead{hongjian.sun@durham.ac.uk}
\author[label3]{Iv\'{a}n Castro}
\author[label4]{and Jing Jiang}

 \cortext[cor1]{Corresponding author.} 

\affiliation[label1]{organization={Department of Engineering, Durham University},
            city={Durham},
            country={UK}}
            
\affiliation[label2]{organization={Department of Computer Sciences, Durham University},
            city={Durham},
            country={UK}}
\affiliation[label3]{organization={Levelise Ltd},
            city={Oxford},
            country={UK}}
\affiliation[label4]{organization={Northumbria University},
            city={Newcastle},
            country={UK} }

\begin{abstract}
By constructing digital twins (DT) of an integrated energy system (IES), one can benefit from DT's predictive capabilities to improve coordinations among various energy converters, hence enhancing energy efficiency, cost savings and carbon emission reduction. This paper is motivated by the fact that practical IESs suffer from multiple uncertainty sources, and complicated surrounding environment. To address this problem, a novel DT-based day-ahead scheduling method is proposed. The physical IES is modelled as a multi-vector energy system in its virtual space that interacts with the physical IES to manipulate its operations. A deep neural network is trained to make statistical cost-saving scheduling by learning from both historical forecasting errors and day-ahead forecasts. Case studies of IESs show that the proposed DT-based method is able to reduce the operating cost of IES by 63.5\%, comparing to the existing forecast-based scheduling methods.
 It is also found that both electric vehicles and thermal energy storages play proactive roles in the proposed method, highlighting their importance in future energy system integration and decarbonisation.
\end{abstract}


\end{frontmatter}


\section{Introduction}
Besides meeting increasing energy demands,  integrated energy systems (IESs) are playing a vital role in reducing carbon emissions for tackling climate crisis \cite{skarvelis2016multiple}. Efforts have been devoted to decarbonising both energy supplies and energy consumption, e.g., promotion of  renewable energy sources, and adoption of electric vehicles (EVs) to replace fossil-fuelled vehicles.  
Versatile energy converters, e.g., combined heat and power (CHP) generation, show great potentials in further improving energy efficiency by jointly meeting multiple energy demands \cite{wang2019optimal}. 
This has stimulated a multi-vector energy system (MVES) modelling method that characterises the operations of multiple energy converters to better coordinate both supply and demand in multi-vector energy forms \cite{moeini2013multiagent}. 

Recently a new concept called digital twins (DT) has received much attention in the literature of IES \cite{en14030774}. With this DT concept, a MVES model can be regarded as a virtual replica of a physical IES \cite{8901113}. One key feature of DT technologies is the interaction between the virtual replica and its physical twin (real-world IES), as well as with external environment. The twinning process between a virtual replica and its physical twin can be achieved by using advanced communication technologies, such as Internet of Things and 5G technologies.  The real-time information of physical system status and operations can be continuously collected and fed to the MVES model, which then simulates, optimises and predicts the future status of its physical twin. These predictions or optimised configurations can be sent to physical devices. Hence, the physical counterparts of DTs, e.g., energy converters and renewable energy sources, could be better coordinated and managed, leading to the reduction of operating cost for IESs. 

Through the use of DT technologies, IESs could not only participate in wholesale energy market, but also provide ancillary services, such as balancing services \cite{UK_E_Data_Lib}. Moreover, by integrating renewable energy sources and flexible loads, the DT can help improve energy efficiency, leading to more cost-savings for both IES owners and energy consumers \cite{xiang2020cost}.  
However, there are multiple uncertainty sources in such a system \cite{ErReport} that might compromise DT's cost-saving potential \cite{xiang2020cost}. Besides the intermittency of renewable energy sources, forecasts could introduce errors that result in uncertainties of system operations \cite{ge2019flexibility}. 
{In traditional IES operation scheduling studies, the uncertainty is usually addressed by providing conserved performance with worst-case assumptions, or simplifying the distribution of the uncertainty sources. For example, in \cite{liu2018day}, the formulated IES integrates with renewable energy, where the renewable energy forecasting errors are considered as uncertainty sources. The distribution of the forecasting errors is mapped to intervals with fixed proportion method, whose performance might depend on the adopted intervals. In \cite{salkuti2019day}, Salkuti proposed a multi-objective optimization based method for the day-ahead scheduling problem in the IES, where the system uncertainty is due to the intermittent renewable energy generation. The uncertainty is addressed by assuming that the distributions of the uncertainty sources are known in prior, which might not be valid if the uncertainty sources are of unknown patterns and unknown distributions in a practical environment.}  
{In the  literature, an economic reserve scheduling problem was formulated for an IES with both electricity and natural gas \cite{liu2018day}, where the renewable energy forecasting error was accounted by an interval with fixed proportion method.
Salkuti \cite{salkuti2019day} studied a multi-objective optimization based method for the day-ahead scheduling of an IES with thermal, wind, and solar energy. In this work, renewable energy generation was assumed to follow a prior probability distribution. But such a strong assumption may not be valid in a practical environment where obtaining priors would be challenging. 
}

{Recently, machine learning based methods are providing promising solutions to the uncertainty problem in IESs. Comparing to traditional analytical based methods, machine learning based methods have the potential to learn from the data to mitigate the issues caused by introducing inappropriate simplifications or assumptions to the IESs. There have been extensive studies using the predictions from the machine learning algorithms to improve forecasting accuracy. For example, Fan {\it et al.} \cite{fan2019assessment} used recurrent neural networks to predict the building's short-term energy demand, while Theocharides {\it et al.} \cite{THEOCHARIDES2020115023} studied a data-driven method to forecast the hourly-averaged day-ahead photovoltaic power generation. There are also works that use machine learning algorithms to calibrate the energy forecasting errors in IESs, for example, Zhu {\it et al.} proposed an approximate Bayesian computation based method to calibrate the building energy models in \cite{ZHU2020115025}. A detailed review of the recent advances of using machine learning for building load prediction can be found in \cite{ZHANG2021116452}.}
{To address the uncertainty problem in IESs, machine learning based methods were extensively studied. 
Fan {\it et al.} \cite{fan2019assessment} used recurrent neural networks to predict the building's short-term energy demand. 
In \cite{THEOCHARIDES2020115023}, Theocharides {\it et al.} studied a data-driven method to forecast the hourly-averaged day-ahead photovoltaic power generation. 
Zhu {\it et al.} \cite{ZHU2020115025} studied an approximate Bayesian computation based machine learning method for calibrating the building energy models to address the energy forecasting errors. 
To understand the full literature of this research area, please refer to a classic work \cite{ZHANG2021116452} by Zhang {\it et al.} who summarised recent advances of using machine learning for building load prediction. 
Moreover, there have been several successful attempts in addressing the day-ahead scheduling problem. 
In \cite{teo2018near}, Teo {\it et al.} proposed an extreme learning machine method for the day-ahead scheduling and real-time dispatching schemes by training historical data to make accurate forecasts. 
In \cite{zhou2019deep}, Zhou {\it et al.} used long short-term memory neural networks to make probabilistic forecasts on load demands and wind energy generations, and integrated them into the look-ahead dispatching schemes. 
Zhou {\it et al.} \cite{zhang2020deep} further studied the day-ahead scheduling problem for forecasting the electricity price  by using a recurrent deep learning method. The performance of this method was verified against real-world data in the New England electricity market.  While aforesaid attempts were made to address single source of uncertainties, there are still some serious limitations when the  day-ahead IES scheduling comes to real-life deployments where multiple uncertainty sources co-exist, which is the main focus of this paper. }

{Besides the literature focusing on improving the prediction accuracy, there are also successful attempts in integrating the machine learning based forecasts into the IES operation scheduling problems. In \cite{teo2018near}, Teo {\it et al.} proposed an extreme learning machine method for the day-ahead scheduling and real-time dispatching schemes by training historical data to make accurate forecasts. 
In \cite{zhou2019deep}, Zhou {\it et al.} used long short-term memory neural networks to make probabilistic forecasts on load demands and wind energy generations, and integrated them into the look-ahead dispatching schemes. 
Zhou {\it et al.} \cite{zhang2020deep} further studied the day-ahead scheduling problem for forecasting the electricity price  by using a recurrent deep learning method. The performance of this method was verified against real-world data in the New England electricity market. However, the aforesaid attempts were made to address a single source of uncertainties, while it leaves an open research challenge to addressing the real-life IESs where multiple uncertainty sources co-exist. This is also the main focus of this paper.}

{With the advent of DT, there is also a novel and promising trend in addressing the IES operation scheduling problem by the integration of DT and machine learning algorithms. In \cite{Agostinelli2021}, Agostinelli {\it et al.} exploited  the machine learning algorithms as part of the co-simulators in the DT models, which helps to improve the building energy management by jointly considering the energy efficiency, internal comfort, and climate conditions.    
Edward {\it et al.} \cite{Edward2020DT} proposed a versatile energy management tool based on DT, which utilized machine learning modules to forecast the energy system demands, coordinated the control of multi-vector smart energy systems. Different from \cite{Agostinelli2021} and \cite{Edward2020DT}, this work did not make local forecasts of the demands or generations, but instead the machine learning technique was used to directly address the IES scheduling problem with multiple uncertainties, whose outputs were the day-ahead scheduling values.}

To fill these research gaps, this paper presents a novel deep learning and DT based IES scheduling method. In contrast to the literature, new contributions of this paper are: 
\begin{enumerate}
\item This deep learning and DT based IES scheduling method is designed to address the problem of multiple uncertainty sources. It provides a more practical solution for operating the real-life IESs embedded with multiple uncertainty sources. 
\item Different from existing work which focus on improving the accuracy of predictions,  the proposed IES scheduling method is designed for enabling real-time intervention and prevention of IESs by taking advantages of DT technologies. Ultimate performance of a physical IES is directly linked to and addressed by its virtual replica's predictive capability. 
\item  A data augmentation based deep learning method is proposed to reduce the MVES's long-term operating cost by learning from both historical forecasting errors and day-ahead forecasts.
\end{enumerate}

The remaining parts of this paper are organised as follows. The DT model of the IES is given in Section \ref{Section Methodology}, and the proposed DT based IES scheduling method is introduced in Section \ref{section proposed}. Case studies with real-world U.K. data are performed in Section \ref{section cases}. Finally the conclusions are presented in Section \ref{section conclusions}. 

\section{System Model}
\label{Section Methodology}
\begin{figure*}[!ht]
\centering
\includegraphics[width= \FigSize , trim={0cm 0.0cm 0.0cm 0.0cm},clip]{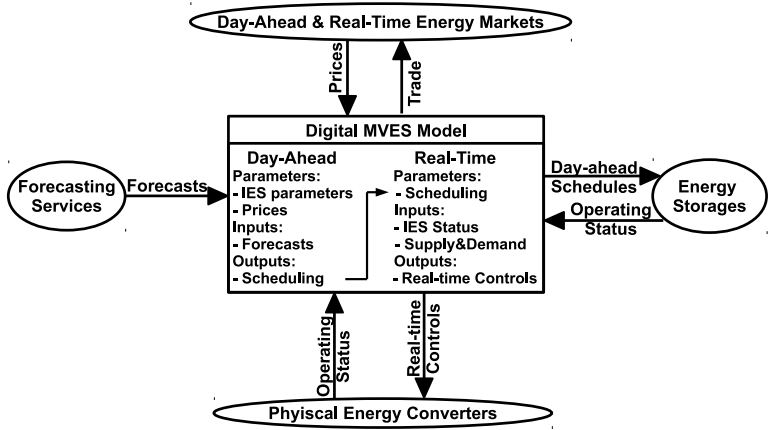}
\caption{{The data flow of a typical IES, where its virtual replica - MVES, and physical IES devices form a DT system to interact with external environment.}}
\label{fig data flow}
\end{figure*}

The DT model consists of energy flows and data flows, which are given in Fig. \ref{fig system setup} and Fig. \ref{fig data flow}. The IES supplies both electricity loads and heat loads with energy sourcing from the electricity, the renewable energy and the natural gas. The IES components are aggregated according to their energy forms, which is to facilitate the DT operation. Please kindly note that since all IES components are studied in an aggregated manner, it is expected that for each type of IES component, there will be multiple of them in numbers, e.g., wind energy could represent a wind farm, while the electricity load could be an aggregated electricity demand from thousands of buildings. Various energy converters are considered, including electrical transformers, natural gas boilers, and CHP systems (from natural gas to electricity and heat).  
Thermal energy storages (TESs) and EVs are also considered, which can charge and discharge energy as scheduled by the MVES.

The DT integration with the real-time operation of the IES is illustrated in Fig. \ref{fig data flow}. Specifically, the virtual replica MVES coordinates all the data flow within the physical IES and among the physical IES devices. During the day-ahead scheduling phase, the MVES makes day-ahead scheduling decisions for the IES devices, including the TESs and EVs, which are based on the forecasts, energy market prices, and the TESs and EVs status. Then the MVES coordinates the IES to make purchases from the day-ahead energy markets, and stores the day-ahead scheduling plans for the next-day operations. During the real-time operation phase, the MVES commits the day-ahead scheduling and monitors the operating status within the IES, while it coordinates the physical energy converters to meet real-time energy demands. The day-ahead scheduling and real-time coordination between the virtual replica MVES and the physical IES devices will be detailed in the following subsections.

\subsection{Multi-vector Energy System Model}
\begin{figure*}[!ht]
\centering
\includegraphics[width= \FigSize , trim={0cm 0.0cm 7cm 0.0cm},clip]{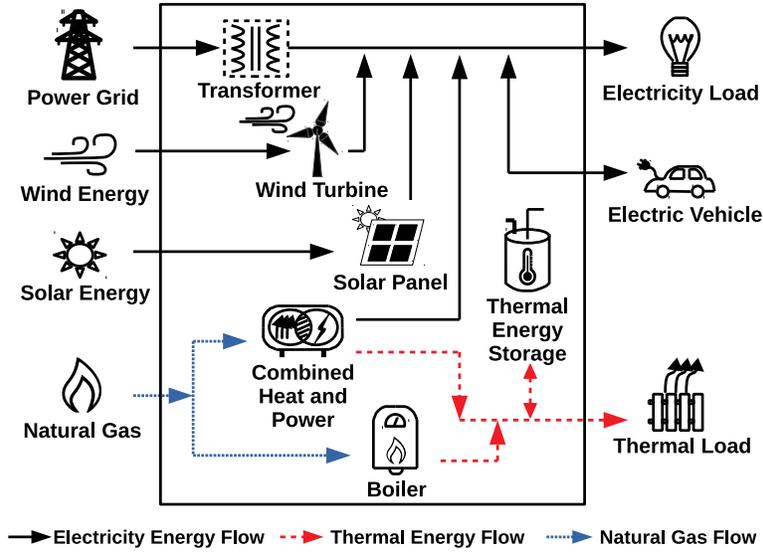}
\caption{The energy flow of a typical IES, which consists of transformers, wind turbines, solar panels,  CHP systems, TESs, boilers and EVs.}
\label{fig system setup}
\end{figure*}

Let $T$ denote the total number of scheduling slots in one day, where time intervals throughout the day have the same length (e.g., 1 hour). At each scheduling slot $t \in [1, T]$, let $S_\text{E}^t$ and $S_\text{G}^t$ denote the imported electricity and natural gas, respectively. 
Within the IES, there are $K_{\text{W}}$ wind turbines, $K_{\text{PV}}$ solar panels, $K_{\text{EV}}$ EVs and $K_{\text{TES}}$ TES systems, whose energy flows at scheduling slot $t$ are denoted as $S_\text{W}^{k, t}$ with $k=1,\dots, K_\text{W}$, $S_\text{PV}^{k, t}$ with $k=1,\dots,K_{\text{PV}}$, $S_\text{EV}^{k, t}$ with $k=1,\dots,K_\text{EV}$ and $S_\text{TES}^{k, t}$ with $k=1,\dots,K_\text{TES}$, respectively. The sign of the energy flow is positive if the energy is injected into the IES, or negative if the energy is sourced from the IES. 
In the considered system in Fig. \ref{fig system setup}, the elements of $\mathbf{v^t} = \{v_\text{CHP}^t, v_\text{B}^t\}$ are the dispatching factors of the CHP and the boiler, respectively. 
The converters are subject to physical constraints during the operations, which are detailed as follows:

\begin{itemize}
\item \textbf{The Electricity Load Constraint:} \\
The electricity supplied by the IES should meet the electricity load demands $L_\text{E}^t$ at each scheduling slot, which can be given by:
\begin{equation}
\label{constraints electricity load}
\eta_\text{TF} S_\text{E}^t  + v_\text{CHP}^t \eta_\text{CHP}^{\text{E}} S_\text{G}^t + \sum_{k=1}^{K_\text{W}} S_\text{W}^{k,t} + \sum_{k=1}^{K_\text{PV}} S_\text{PV}^{k,t} + \sum_{k=1}^{K_\text{EV}} S_\text{EV}^{k,t} = L_\text{E}^t,
\end{equation}
where $\eta_\text{TF}$ and {$\eta_\text{CHP}^{\text{E}}$}{$\eta_\text{CHP}^{E}$} are the transformer's energy efficiency and the CHP's electricity generation efficiency, respectively. 
\item \textbf{The Heat Load Constraint:} \\
The thermal energy supplied by the IES should meet the heat load demands $L_\text{H}^t$ at each scheduling slot, which can be given by:
\begin{equation}
\label{constraints heat load}
v_\text{CHP}^t \eta_\text{CHP}^\text{H} S_\text{G}^t + v_\text{B}^t \eta_\text{B} S_\text{G}^t + \sum_{k=1}^{K_\text{TES}} S_\text{TES}^{k,t} = L_\text{H}^t,
\end{equation}
where $\eta_\text{CHP}^\text{H}$ is the thermal recovery efficiency of the CHPs, and $\eta_\text{B}$ is the energy efficiency of the boilers. 
\item \textbf{The Energy Flow Constraints of Transformers, Natural Gas, Wind Turbines, Solar Panels, CHPs and Boilers:} \\
If we denote the maximum energy flow of transformers, natural gas, wind turbines, solar panels, CHPs and boilers at a scheduling slot as $S_\text{TF}^\text{MAX}$, $S_\text{G}^\text{MAX}$, $S_\text{W}^\text{MAX}$, $S_\text{PV}^\text{MAX}$, $S_\text{CHP}^\text{MAX}$ and $S_\text{B}^\text{MAX}$, respectively, then the scheduled energy flows should satisfy corresponding minimal and maximal energy flow constraints at each scheduling slot, which can be formulated as follows:
\begin{align}
\label{constraints transformer power}
&0 \leq S_\text{E}^t \leq S_\text{TF}^\text{MAX}, \\
\label{constraints natural gas}
&0 \leq S_\text{G}^t \leq S_\text{G}^\text{MAX}, \\
\label{constraints wind turbine power}
&0 \leq S_\text{W}^{k, t} \leq S_\text{W}^\text{MAX}, k = 1, \dots, K_\text{W},\\
\label{constraints PV power}
&0 \leq S_\text{PV}^{k, t} \leq S_\text{PV}^\text{MAX}, k = 1,\dots, K_\text{PV},\\
\label{constraints CHP power}
&0 \leq v_\text{CHP}^t S_\text{G}^{t} \leq S_\text{CHP}^\text{MAX},\\
\label{constraints Boiler power}
&0 \leq v_\text{B}^t S_\text{B}^{t} \leq S_\text{B}^\text{MAX}.
\end{align}
\item \textbf{The EVs' Charging and Discharging Constraints:}\\
A total of $K_\text{EV}$ EVs are considered, where the $k$th EV's scheduled start and end of service time are denoted as $T_\text{EV}^{\text{In},k}$ and $T_\text{EV}^{\text{Out},k}$, respectively. The energy flow observed from the $k$th EV's aspect is denoted as $S_\text{EV}^{k, t}$, which can be: a) positive if discharging, b) negative if charging, and c) 0 if neither charging nor discharging. For notation convenience, the EV's charging state indicator $I_\text{EV}^{\text{Ch},k}$ and discharging state indicator $I_\text{EV}^{\text{DCh},k}$ are defined as follows:
\begin{equation}
\begin{aligned}
\label{eq EV state indicator}
I_\text{EV}^{\text{Ch}, k,t} \triangleq \frac{1-\text{sgn}\{S_\text{EV}^{k, t}\}}{2},
I_\text{EV}^{\text{DCh},k,t} \triangleq \frac{1+\text{sgn}\{S_\text{EV}^{k, t}\}}{2},
\end{aligned}
\end{equation}
where the sign operator $\text{sgn}\{x\}$ returns 1 if $x>0$ and -1 if $x<0$, otherwise it will return 0. Note that due to the charging and discharging efficiency, the energy flow observed from the aspect of MVES is the total energy consumed by this EV, consisting of both the energy charged to or discharged from this EV, and the energy loss due to the charging and discharging operations. Denote the charging and discharging efficiency as $\eta_\text{EV}^\text{Ch}$ and $\eta_\text{EV}^\text{DCh}$, then the consumed energy by the EV can be given as $S_\text{EV}^{k, t} \left( \frac{I_\text{EV}^{\text{Ch}, k,t}}{\eta_\text{EV}^\text{Ch}} + \frac{I_\text{EV}^{\text{DCh}, k,t}}{\eta_\text{EV}^\text{DCh}} \right)$. 
With the indicators $I_\text{EV}^{\text{Ch}, k,t}$ and $I_\text{EV}^{\text{DCh},k,t}$, it can be seen that the energy flow observed from the aspect of MVES is: a) $\frac{S_\text{EV}^{k, t}}{\eta_\text{EV}^\text{Ch}}$, when the EV is charging, b) $\frac{S_\text{EV}^{k, t}}{\eta_\text{EV}^\text{DCh}}$, when the EV is discharging, and c) $0$, when the EV is neither charging nor discharging.
Then the $k$th EV's energy flow constraint at the scheduling slot $t\in[T_\text{EV}^{\text{In},k}, T_\text{EV}^{\text{Out},k}]$ can be given as follows:
\begin{equation} 
\label{constraints EV power}
-S_\text{EV}^\text{Ch, MAX} \leq S_\text{EV}^{k, t} \left( \frac{I_\text{EV}^{\text{Ch}, k,t}}{\eta_\text{EV}^\text{Ch}} + \frac{I_\text{EV}^{\text{DCh}, k,t}}{\eta_\text{EV}^\text{DCh}} \right) \leq S_\text{EV}^\text{DCh, MAX}, 
\end{equation}
where $S_\text{EV}^\text{Ch, MAX}$ and $S_\text{EV}^\text{DCh, MAX}$ are the maximal charging and discharging energy flows within a scheduling slot period. 
In the meantime, the MVES rewards EVs by leasing their available capacities as electricity energy buffers, therefore the EVs' original energy levels should be restored at the end of their service time. This corresponds to a zero net energy flow for the $k$th EV during $[T_\text{EV}^{\text{In},k}, T_\text{EV}^{\text{Out},k}]$, which can be formulated as follows:
\begin{equation}
\label{constraints EV SOC equal}
\sum_{t= T_\text{EV}^{\text{In},k}}^{T_\text{EV}^{\text{Out},k}}  S_\text{EV}^{k,t} = 0.
\end{equation}  
In this paper, we use $\text{SOC}_\text{EV}^{k,T_\text{EV}^{\text{In},k}}$ to represent the $k$th EV's State-of-Charge (SOC) at its start service time $T_\text{EV}^{\text{In},k}$, then the $\text{SOC}_\text{EV}^{k,t}$ for the scheduling slot $t\in [T_\text{EV}^{\text{In},k}, T_\text{EV}^{\text{Out},k}]$ can be given as follows:
\begin{equation}
\label{eq EV SOC}
\begin{aligned}
\text{SOC}_\text{EV}^{k,t} = \text{SOC}_\text{EV}^{k, T_\text{EV}^{\text{In},k}} + \sum_{\tau= T_\text{EV}^{\text{In},k}}^t S_\text{EV}^{k,t}.
\end{aligned}
\end{equation}
It should be noted that for any scheduling slot $t$, the $k$th EV's $\text{SOC}_\text{EV}^{k,t}$ should be bounded by the minimal energy capacity $\text{SOC}_\text{EV}^\text{MIN}$ and the maximal energy capacity $\text{SOC}_\text{EV}^\text{MAX}$, which can be formulated as follows:
\begin{equation}
\label{constraints EV SOC limit}
\begin{aligned}
\text{SOC}_\text{EV}^\text{MIN} \leq \text{SOC}_\text{EV}^{k,t} \leq \text{SOC}_\text{EV}^\text{MAX}. 
\end{aligned}
\end{equation}
\item \textbf{The TES Charging and Discharging Constraints:}\\
The TES is a flexible thermal energy storage, whose model is similar to the EVs as above. The start and end of the service time for the $k$th TES system are denoted as $T_\text{TES}^{\text{In},k}$ and $T_\text{TES}^{\text{Out},k}$, respectively. The TES's charging state indicator $I_\text{TES}^{\text{Ch},k}$ and discharging state indicator $I_\text{TES}^{\text{DCh},k}$ are defined as follows:
\begin{equation}
\begin{aligned}
\label{eq TES state indicator}
I_\text{TES}^{\text{Ch}, k,t} \triangleq \frac{1-\text{sgn}\{S_\text{TES}^{k, t}\}}{2},
I_\text{TES}^{\text{DCh},k,t} \triangleq \frac{1+\text{sgn}\{S_\text{TES}^{k, t}\}}{2}.
\end{aligned}
\end{equation}
The energy flow $S_\text{TES}^{k,t}$ of the $k$th TES at the scheduling slot $t \in [T_\text{TES}^{\text{In},k}, T_\text{TES}^{\text{Out},k}]$ are constrained as follows:
\begin{equation} 
\label{constraints TES power}
-S_\text{TES}^\text{Ch, MAX} \leq S_\text{TES}^{k, t} \left( \frac{I_\text{TES}^{\text{Ch}, k,t}}{\eta_\text{TES}^\text{Ch}} + \frac{I_\text{TES}^{\text{DCh}, k,t}}{\eta_\text{TES}^\text{DCh}} \right) \leq S_\text{TES}^\text{DCh, MAX}, 
\end{equation}
where $S_\text{TES}^\text{Ch, MAX}$ and $S_\text{TES}^\text{DCh, MAX}$ are the maximal charging and discharging energy flow within a scheduling slot period, while $\eta_\text{TES}^\text{DCh}$ and $\eta_\text{TES}^\text{Ch}$ denote the energy efficiency of the TES during discharging and charging. 
In the meantime, the MVES rewards TESs by leasing their available capacities as thermal energy buffers, therefore the TESs' original energy levels should be restored at the end of their service time. This equivalents to a zero net energy flow during $[T_\text{TES}^{\text{In},k}, T_\text{TES}^{\text{Out},k}]$ as follows:
\begin{equation}
\label{constraints TES SOC equal}
\sum_{t= T_\text{TES}^{\text{In},k}}^{T_\text{TES}^{\text{Out},k}} S_\text{TES}^{k,t} = 0.
\end{equation}   
Let $\text{SOC}_\text{TES}^{k,T_\text{TES}^{\text{In},k}}$ denote the $k$th TES's SOC at  $T_\text{TES}^{\text{In},k}$, then the $\text{SOC}_\text{TES}^{k,t}$ for the scheduling slot $t\in [T_\text{TES}^{\text{In},k}, T_\text{TES}^{\text{Out},k}]$ can be given as follows:
\begin{equation}
\label{eq TES SOC}
\begin{aligned}
\text{SOC}_\text{TES}^{k,t} =  \text{SOC}_\text{TES}^{k, T_\text{TES}^{\text{In},k}} +
\sum_{\tau= T_\text{TES}^{\text{In},k}}^t S_\text{TES}^{k,t}.
\end{aligned}
\end{equation}
For any scheduling slot $t$, the TES's $\text{SOC}_\text{TES}^{k,t}$ should be bounded as follows: 
\begin{equation}
\label{constraints TES SOC limit}
\begin{aligned}
\text{SOC}_\text{TES}^\text{MIN} \leq \text{SOC}_\text{TES}^{k,t} \leq \text{SOC}_\text{TES}^\text{MAX},
\end{aligned}
\end{equation}
where $\text{SOC}_\text{TES}^\text{MIN}$ and $\text{SOC}_\text{TES}^\text{MAX}$ are the TES's minimal and maximal energy capacities, respectively.
\end{itemize}

\subsection{Day-ahead Scheduling with Multiple Uncertainty Sources}
\label{subsection real-time operation}
In the problem of day-ahead scheduling with multiple uncertainty sources, the system inputs are the day-ahead forecasts $\mathbf{F}^t \triangleq \{{L}_\text{E}^{t}, {L}_\text{H}^{t}, S_\text{W}^{k,t}, S_\text{PV}^{k,t} \}$, while the outputs are the day-ahead scheduling  $\mathbf{S}_\text{Sch}^t \triangleq \{S_\text{E}^{t}, S_\text{G}^{t}, S_\text{EV}^{k,t}, S_\text{TES}^{k,t}, \mathbf{v}^{t}\}$.
The overall objective is to minimize the total cost $C_\text{All}^t$ given as follows:
\begin{equation}
\label{eq total cost rewrite}
C_\text{All}^t = C_\text{Sch}^t + C_\text{Extra}^t,
\end{equation}
where the scheduling cost $C_\text{Sch}^t$ and the extra cost $C_\text{Extra}^t$ are detailed as follows:
\begin{itemize}
\item \textbf{The {Day-Ahead} Scheduling Cost $C_\text{Sch}^t$}:\\
For the scheduling slot $t$ in the day-ahead markets, the electricity price and the natural gas price are denoted as $C_\text{E}^{\text{DA}, t}$ and $C_\text{G}^{\text{DA}, t}$, respectively. 
Meantime, the reward price for the wind energy, solar energy, EVs and TESs are fixed all day and denoted as $C_\text{W}$, $C_\text{PV}$, $C_\text{EV}$ and $C_\text{TES}$, respectively. 
Therefore the total day-ahead scheduling cost $C_\text{Sch}^t$ can be formulated as follows:
\begin{equation}
\label{eq C sch}
\begin{aligned}
& C_\text{Sch}^t =  C_\text{E}^{\text{DA}, t} S_\text{E}^{t} + C_\text{G}^{\text{DA}, t}S_\text{G}^{t} - \sum_{k=1}^{K_\text{TES}} C_\text{TES} |S_\text{TES}^{k,t}| \\& -\sum_{k=1}^{K_\text{EV}} C_\text{EV} |S_\text{EV}^{k,t}| - \sum_{k=1}^{K_\text{PV}} C_\text{PV} S_\text{PV}^{k,t}  -  \sum_{k=1}^{K_\text{W}} C_\text{W} S_\text{W}^{k,t} ,
\end{aligned}
\end{equation}
{where the EVs and TESs are rewarded based on their service to the DT system, which is quantified by their absolute energy flows.}{where the rewards to the EVs and TESs are based on their absolute energy flows}.

\item \textbf{The {Real-Time Operation} Extra Cost $C_\text{Extra}^t$}:\\
During the real-time operation, the IES follows the day-ahead scheduling $\mathbf{S}_\text{Sch}^t$ to operate the CHPs, EVs and TESs. Due to the forecasting errors, the actual electricity load demands $\tilde{L}_\text{E}^t$, the actual heat load demands $\tilde{L}_\text{H}^t$, the actual wind energy generation $\tilde{S}_\text{W}^t$ and the actual solar energy generation $\tilde{S}_\text{PV}^t$ can be different from their forecasting values, which results in the  mismatch between real-time demands and supplies. The MVES addresses this real-time mismatch via the transformer and the boiler as follows: a) in the case of insufficient electricity energy, the MVES will purchase the insufficient amount of electricity {in the real-time energy markets} with a price $C_\text{E}^+$, and inject to the MVES via the transformer, b) in the case of redundant electricity energy, the MVES will refund {or trade-in} the redundant amount of electricity {in the real-time energy markets} with a price $C_\text{E}^-$, and reduce the injected electricity from transformer, c) in the case of insufficient thermal energy, the MVES will purchase the insufficient amount of natural gas {in the real-time energy markets} with a price $C_\text{G}^+$, and inject to the boiler to compensate the insufficient thermal energy, and d) in the case of redundant thermal energy, the MVES will refund {or trade-in} the redundant amount of natural gas {in the real-time energy markets} with a price $C_\text{G}^-$, and reduce the consumed natural gas by the boiler. {Note that during the real-time operation, the charging and discharging of the EVs and TESs are following the day-ahead scheduling decisions. This is because the charging and discharging control of the EVs and TESs not only depends on their historical operations, but also impacts their future operations. Therefore the real-time adjustment might not be optimal from the view of the whole day operation cost reduction.} Then the extra cost $C_\text{Extra}^t$ can be given as follows:
\begin{equation}
\label{eq total cost}
C_\text{Extra}^t = C_\text{\text{Extra},E}^{t} + C_\text{\text{Extra},G}^{t},
\end{equation}
where the extra cost for electricity $C_\text{\text{Extra},E}^{t}$ and natural gas $C_\text{\text{Extra},G}^{t}$ can be given as follows:
\begin{equation}
\begin{aligned}
\label{eq cost punish e and g}
C_\text{\text{Extra},E}^{t} = \left\{\begin{matrix}
C_\text{E}^+ \Delta S_\text{E}^{t}, & \Delta S_\text{E}^{t} \geq 0,\\ 
(C_\text{E}^{\text{DA}, t} - C_\text{E}^- ) \Delta S_\text{E}^{t}, &\Delta S_\text{E}^{t} <  0.
\end{matrix}\right. \\
C_\text{\text{Extra},G}^{t} = \left\{\begin{matrix}
C_\text{G}^+ \Delta S_\text{G}^{t}, &  \Delta S_\text{G}^{t}  \geq 0,\\ 
(C_\text{G}^{\text{DA}, t} - C_\text{G}^-)  \Delta S_\text{G}^{t}, &  \Delta S_\text{G}^{t}  <  0.
\end{matrix}\right. 
\end{aligned}
\end{equation}
where $\Delta S_\text{E}^{t}$ and $\Delta S_\text{G}^{t}$ denote the mismatched electricity and natural gas, which are calculated as follows:
\begin{equation}
\begin{aligned}
\Delta S_\text{E}^{t} = & \frac{1}{\eta_\text{TF}}\Big(\tilde{L}_\text{E}^t - \eta_\text{TF} S_\text{E}^t  - v_\text{CHP}^t \eta_\text{CHP}^{\text{E}} S_\text{G}^t  \\
& -   \sum_{k=1}^{K_\text{W}} \tilde{S}_\text{W}^{k,t} -\sum_{k=1}^{K_\text{PV}} \tilde{S}_\text{PV}^{k,t} - \sum_{k=1}^{K_\text{EV}} S_\text{EV}^{k,t}\Big) ,
\end{aligned}
\end{equation}
\begin{equation}
\Delta S_\text{G}^{t} = \frac{1}{\eta_\text{B}} \Big( \tilde{L}_\text{H}^t - v_\text{CHP}^t \eta_\text{CHP}^\text{H} S_\text{G}^t - v_\text{B}^t \eta_\text{B} S_\text{G}^t - \sum_{k=1}^{K_\text{TES}} S_\text{TES}^{k,t} \Big). 
\end{equation}
\end{itemize}

If with no forecasting errors, the IES operator can avoid any extra cost $C_\text{Extra}^t$, because both $\Delta S_\text{E}^{t}$ and $\Delta S_\text{G}^{t}$ will be zero. However, during the day-ahead scheduling, the exact forecasting errors $\boldsymbol{\delta}^t$ cannot be known, which means the extra costs $C_\text{Extra}^t$ are random and unavoidable. With the consideration of the day-ahead forecasts and the potential forecasting errors $\boldsymbol{\delta}^t$, it is potential to make more cost-effective day-ahead scheduling, comparing to schedules based solely on the day-ahead forecasts. {Please also note that the forecast errors are random in nature, while the reported forecasting performances are usually referring to its average performance in the long-term, e.g., the historical U.K. dataset is referring to a period of 1 year. However, the uncertainty introduced by these forecasting errors are impacting the IES system hourly (or less then 1 hour depending on the forecast resolutions). These time-varying forecasting errors can have unknown patterns, which are also not possible to be bounded in the real-world (e.g., due to unexpected conditions like bad weather). Therefore the averaged forecasting performance (e.g., the yearly averaged forecasting errors) only reflects part of the uncertainty in the practical IES. By addressing the uncertainty with the consideration of their time-varying patterns, it is potential to further reduce the extra costs induced by these uncertainties.} 

The day-ahead scheduling objective is formulated as the minimisation of the mathematical expectation of the actual total cost $C_\text{All}^t$ against potential forecasting errors $\boldsymbol{\delta}^t$ {for the whole 24 hour period} as follows:
\begin{align}
\label{eq objective function uncertainty}
&\mathbb{P}_1:  & \min_{\mathbf{S}_\text{Sch}^t} &  \quad \mathbb{E}_{\boldsymbol{\delta}^t} \left\lbrace \sum_{t=1}^{T} C_\text{All}^t \right\rbrace \\
& {} & \text{ s.t. } & \eqref{constraints electricity load}-\eqref{constraints Boiler power}, \eqref{constraints EV power}-\eqref{constraints EV SOC equal}, \notag \\
& {} & {} & \eqref{constraints EV SOC limit}, \eqref{constraints TES power}-\eqref{constraints TES SOC equal} \text{ and } \eqref{constraints TES SOC limit} \notag,
\end{align}
where  $\mathbb{E}_{x}\{y(x)\} {=} \int_{- \infty}^{\infty} y(x) p(x) dx$ is the mathematical expectation operation and $p(x)$ is the probability distribution function (PDF) of $x$. 

The formulation of $\mathbb{P}_1$ can be further interpreted as follows. With the given day-ahead forecasts $\mathbf{F}^t$, the MVES is to make a day-ahead scheduling $\mathbf{S}_\text{Sch}^t$, such that it can: a) have the lowest cost in a statistical manner against the potential forecasting errors $\boldsymbol{\delta}^t$, and b) meet the physical constraints as specified in \eqref{constraints EV SOC limit}, \eqref{constraints TES power}-\eqref{constraints TES SOC equal} \text{ and } \eqref{constraints TES SOC limit}. This makes $\mathbb{P}_1$ a very challenging problem, because each forecast $\mathbf{F}^t$ could correspond to countless actual values $\mathbf{\tilde{F}}^{t}$  due to the random forecasting errors $\boldsymbol{\delta}^t$, where any day-ahead scheduling could be saving costs for some actual values $\mathbf{\tilde{F}}^{t}$, while increasing costs in other cases at the same time. Here $\mathbf{\tilde{F}}^{t} \triangleq \{\tilde{L}_\text{E}^{t}, \tilde{L}_\text{H}^{t}, \tilde{S}_\text{W}^{k,t}, \tilde{S}_\text{PV}^{k,t}\}$, whose elements are the actual electricity load demand, the actual heat load demand, the actual wind energy generation, and the actual solar energy generation, respectively. In this paper, this challenge is addressed via a deep learning embedded scheduling scheme, which is detailed in the next section.
\section{Deep Learning Embedded Integrated Energy System Scheduling Scheme}
\label{section proposed}
To address the challenge of day-ahead scheduling under multiple uncertainties for IESs, we propose a novel deep learning embedded scheduling scheme, which learns the statistical optimal day-ahead scheduling from the historical forecasts and forecasting errors.
\begin{figure*}[!htbp]
\centering
\includegraphics[width=\FigSize, trim={6.5cm 0.0cm 8cm 0.0cm},clip]{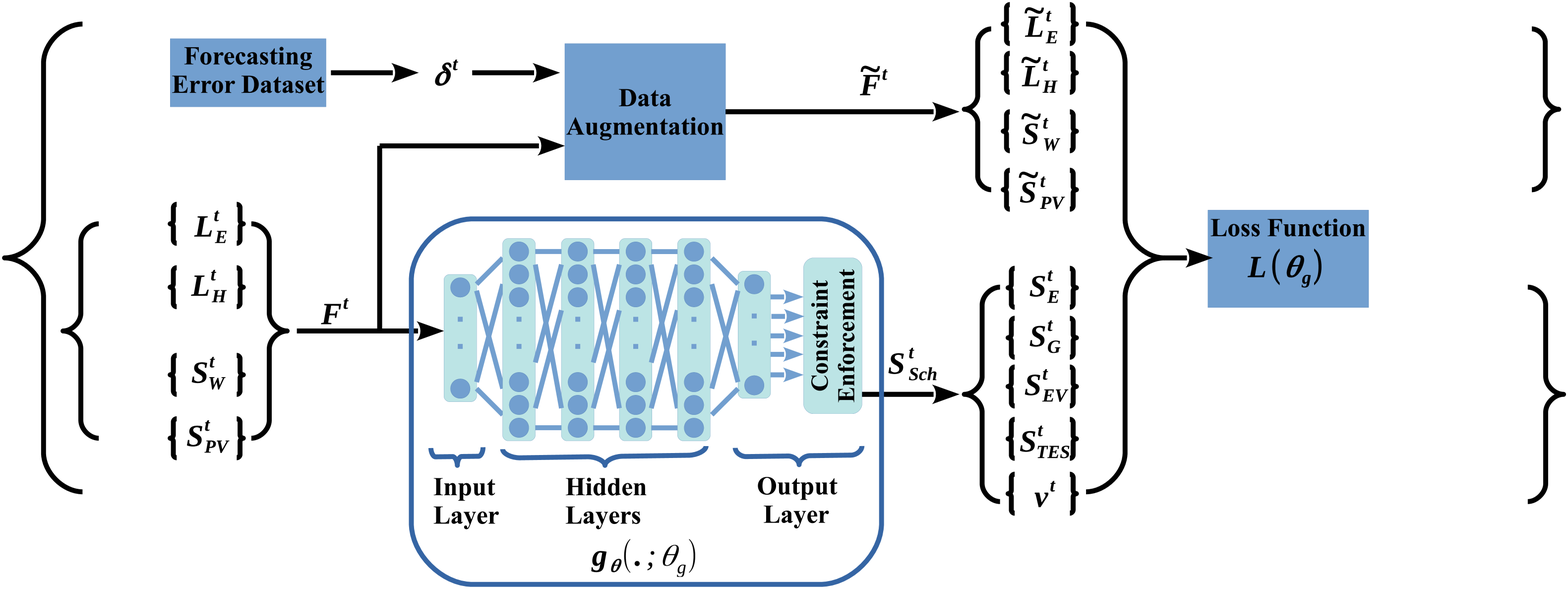}
\caption{The proposed deep learning based training approach for deep
neural network $g_{\boldsymbol{\theta}}(\cdot; \boldsymbol{\theta}_g)$, which takes the day-ahead forecasts $\mathbf{F}^t \triangleq \{{L}_\text{E}^{t}, {L}_\text{H}^{t}, S_\text{W}^{k,t}, S_\text{PV}^{k,t} \}$ as inputs and gives the statistical day-ahead scheduling $\mathbf{S}_\text{Sch}^t \triangleq \{S_\text{E}^{t}, S_\text{G}^{t}, S_\text{EV}^{k,t}, S_\text{TES}^{k,t}, \mathbf{v}^{t}\}$ as outputs.}
\label{fig DNN structure}
\end{figure*}  
\subsection{A Deep Learning Based Approach for the Day-Ahead Scheduling under Multiple Uncertainties}
The proposed deep learning based approach for the day-ahead scheduling under multiple uncertainties is illustrated in Fig. \ref{fig DNN structure}. It consists of a deep
neural network $g_{\boldsymbol{\theta}}(\cdot; \boldsymbol{\theta}_g)$ with a parameter set $\boldsymbol{\theta}_g$, whose inputs are the day-ahead forecasts $\mathbf{F}^t$ and outputs are the day-ahead scheduling $\mathbf{S}_\text{Sch}^t$ as follows:
\begin{equation}
\label{eq g theta}
\mathbf{S}_\text{Sch}^t = g_{\boldsymbol{\theta}}(\mathbf{F}^t;\boldsymbol{\theta}_g),
\end{equation}
where the parameter set $\boldsymbol{\theta}_g$ is trained following Fig. \ref{fig DNN structure}. Once the training procedure is completed, the deep
neural network (DNN) $g_{\boldsymbol{\theta}}(\cdot; \boldsymbol{\theta}_g)$ itself is sufficient to commit the day-ahead scheduling task. The actual total cost $C_\text{ALL}^t$ in \eqref{eq total cost rewrite} can be rewritten by the outputs of the DNN $g_{\boldsymbol{\theta}}(\cdot; \boldsymbol{\theta}_g)$ as {$C_\text{ALL}^t(\mathbf{\tilde{F}}^t, g_{\boldsymbol{\theta}}(\mathbf{{F}^t}; \boldsymbol{\theta}_g))$}{$C_\text{ALL}^t(\mathbf{\tilde{F}}^t, g(\mathbf{{F}^t}; \boldsymbol{\theta}_g))$}. In this way, the original problem $\mathbb{P}_1$ is transformed to a deep learning problem in finding the optimal parameter set $\boldsymbol{\theta}_g$.

\subsection{Enforcing the Physical Constraints}
\label{subsection Enforcing the Physical Constraints}
It is challenging to exploit the DNN to address general problems with physical constraints. This is because in the context of energy systems, each DNN output is with physical interpretations, e.g., the energy amounts or the control ratios. Based on their physical meanings in the real world, they could be involved in multiple physical constraints at the same time. This paper addresses all physical constraints via: a) the ``Constraint Enforcement'' module at the DNN's output layer as illustrated in Fig. \ref{fig DNN structure}, and b) the penalty terms in the loss function design, which are detailed as follows:

\begin{itemize}
\item  The energy flow constraints in \eqref{constraints transformer power}-\eqref{constraints CHP power}, \eqref{constraints EV power} and \eqref{constraints TES power} are enforced by scaling the energy flow values to the range specified by the minimal and maximal limits. Specifically, this is implemented by: a) transforming energy flow output of the transformer, wind turbines, PVs, and CHPs into the range of $[0,1]$, and then scaling them by their corresponding minimal and maximal limits, and b) transforming each energy flow output of EVs and TESs into the range of $[-1,1]$, and then scaling them by their corresponding minimal and maximal limits. 

\item The daily net energy flow constraints of EVs and TESs in \eqref{constraints EV SOC equal} and \eqref{constraints TES SOC equal} are enforced by subtracting their corresponding daily average values. Specifically, this is implemented by: a) firstly calculating the daily average energy flows of each EV and TES, b) substracting each energy flow value by the corresponding daily average energy flow value, and c) scaling each energy flow value to meet the energy flow constraints \eqref{constraints EV power} and \eqref{constraints TES power}. 
\item The EVs' SOC and TESs' SOC related constraints in \eqref{constraints EV SOC limit} and \eqref{constraints TES SOC limit} are regarded as penalty terms, which is formulated as follows:
\begin{equation}
\label{eq penalty function}
\begin{aligned}
&C_\text{Penalty}^t \Big(g_{\boldsymbol{\theta}}(\mathbf{F}^t; \boldsymbol{\theta}_g)\Big)= \\
&\underbrace{\max \{ \text{SOC}_\text{EV}^\text{MIN} - \text{SOC}_\text{EV}^{k,t}, \text{SOC}_\text{EV}^{k,t} - \text{SOC}_\text{EV}^\text{MAX}, 0\}}_{\text{EVs' SOC Constraints}} +\\
& \underbrace{\max \{ \text{SOC}_\text{TES}^\text{MIN} - \text{SOC}_\text{TES}^{k,t}, \text{SOC}_\text{TES}^{k,t} - \text{SOC}_\text{TES}^\text{MAX}, 0\}}_{\text{TESs' SOC Constraints}}.
\end{aligned}
\end{equation}

\item The loss function $\mathcal{L} (\boldsymbol{\theta}_g)$ in Fig. \ref{fig DNN structure} is defined as follows:
\begin{equation}
\label{eq loss function g}
\begin{aligned}
\mathcal{L} (\boldsymbol{\theta}_g) = & \mathbb{E}_{\mathbf{F}^{t}, \boldsymbol{\delta^t}} \Bigg\{ \sum_{t=1}^T C_\text{ALL}^t\left(\mathbf{\tilde{F}}^t,  g_{\boldsymbol{\theta}}(\mathbf{F}^t; {\boldsymbol{\theta}_g}) \right) \\
& + \lambda \sum_{t=1}^T C_\text{Penalty}^t \Big( g_{\boldsymbol{\theta}}(\mathbf{F}^t; \boldsymbol{\theta}_g)\Big) \Bigg\},
\end{aligned}
\end{equation}
where the penalty parameter $\lambda$ is a positive scalar. 
\end{itemize}

Then the day-ahead scheduling under multiple uncertainties can be transformed to the equivalent unconstrained deep learning problem as follows:
\begin{equation}
\min_{\boldsymbol{\theta}_g} \mathcal{L}(\boldsymbol{\theta}_g),
\end{equation} 
which is trained in an unsupervised manner in this paper. {Note that here the unsupervised training method is used instead of the supervised training method. This is because, in the studied practical DT system, the multiple uncertainty sources can have unknown statistical properties, which makes it impractical to calculate the optimal solutions with each DT system input required by the supervised training method. Instead, the unsupervised training method is applied, which trains the NN to optimize the DT end performance, by learning from the underlying relationship between the DT system input, multiple uncertainty sources, and the DT system end performance.}

Note that in $\mathbb{P}_1$, the expectation is with regard to $\boldsymbol{\delta^t}$, i.e., the solution is bind to the specific day-ahead forecast $\mathbf{F}^t$, while it requires a totally new search for other inputs. 
Different from that, the mathematical expectation of $\mathcal{L} (\boldsymbol{\theta}_g)$ in \eqref{eq loss function g} is with respect to both $\boldsymbol{\delta^t}$ and $\mathbf{F}^t$, i.e., the solutions for different $\mathbf{F}^t$ can be obtained via the same trained DNN $g_{\boldsymbol{\theta}}(\cdot; \boldsymbol{\theta}_g)$. 

\subsection{Data Augmentation based Approach for Statistical Training}
During the training procedure, the outputs of $g_{\boldsymbol{\theta}}(\cdot; \boldsymbol{\theta}_g)$ need to be evaluated in a statistical way, which is achieved by calculating the loss function $\mathcal{L} (\boldsymbol{\theta}_g)$ against all potential actual values under random forecasting errors.
In this paper, we study the day-ahead scheduling problem with multiple uncertainties $\boldsymbol{\delta}^t = \{\delta_\text{E}^t, \delta_\text{H}^t, \delta_\text{W}^t, \delta_\text{PV}^t\}$, including the electricity load forecasting error $\delta_\text{E}^t=\frac{\tilde{L}_\text{E}^t - L_\text{E}^t}{L_\text{E}^t}$, the heat load forecasting error $\delta_\text{H}^t = \frac{\tilde{L}_\text{H}^t - L_\text{H}^t}{L_\text{H}^t}$, the wind energy forecasting error $\delta_\text{W}^t = \frac{\tilde{S}_\text{W}^t - S_\text{W}^t}{S_\text{W}^t}$, and the solar energy forecasting error $\delta_\text{PV}^t = \frac{\tilde{S}_\text{PV}^t - S_\text{PV}^t}{S_\text{PV}^t}$. In this way, for each given day-ahead forecast $\mathbf{F}^t$, a set of forecasting errors $\boldsymbol{\delta}^t$ can be randomly selected from historical observations or simulations, which together generate an augmented set of actual values $\mathbf{\tilde{F}}^t$ as illustrated in Fig. \ref{fig DNN structure}.

The data augmentation based approach for the statistical training of $g_{\boldsymbol{\theta}}(\cdot; \boldsymbol{\theta}_g)$ is detailed as follows:
\begin{itemize}
\item For each forecast vectors, a training data set can be generated by historical observations or simulations, which forms the total forecast dataset $\mathbb{F}^t$. During the training procedure, a subset of day-ahead forecasts $\mathbf{F}^t$ is randomly sampled from the total forecast dataset $\mathbb{F}^t$.
\item The forecasting error data set $\Delta^t$ can be generated by historical observations or simulations. 
During the training procedure, a subset of forecasting errors $\boldsymbol{\delta^t}$ is randomly sampled from $\Delta^t$. Then the corresponding potential actual values $\mathbf{\tilde{F}}^{t}$ are calculated via the day-ahead forecasts $\mathbf{F}^t $ and the sampled forecasting errors $\boldsymbol{\delta^t}$, {where $\mathbf{\tilde{F}}^{t} = \{\tilde{F}^t| \tilde{F}^t = (1+\delta^t) F^t, \forall \delta^t \in \boldsymbol{\delta^t}, \forall  F^t \in  \mathbf{F}^t  \}$}.
\item The DNN $g_{\boldsymbol{\theta}}(\cdot; \boldsymbol{\theta}_g)$ takes the day-ahead forecasts $\mathbf{F}^t$ as inputs, and gives the statistical day-ahead scheduling $\mathbf{S}_\text{Sch}^t$ as outputs. 
\item For each statistical day-ahead scheduling $\mathbf{S}_\text{Sch}^t$, it is evaluated against each potential actual values from the augmented data set $\mathbf{\tilde{F}}^{t}$, which forms the loss function $\mathcal{L} (\boldsymbol{\theta}_g)$ in \eqref{eq loss function g}. 
\item For each training step $\tau$, the DNN parameters $\boldsymbol{\theta}_g$ are updated via the Gradient Descent algorithm \cite{sutskever2013importance} as follows:
\begin{equation}
\boldsymbol{\theta}_g^{(\tau)} = \boldsymbol{\theta}_g^{(\tau-1)} - \alpha \nabla_{\boldsymbol{\theta}_g} \mathcal{L}(\boldsymbol{\theta}_g^{(\tau-1)}),
\end{equation}
where $\alpha$ is the learning rate of the training procedure.
\end{itemize}

\begin{table}[!htbp]
\centering
\scriptsize
\caption{Parameters in the Case Studies}
\label{table parameters}
\begin{tabular}{ l l r }
\hline
\textbf{Parameter} & \textbf{Description} & \textbf{Value} \\ \hline
$\eta_\text{TF}$         & Transformer's efficiency  & 0.980 \\ 
$\eta_\text{CHP}^\text{E}$         & CHP's power generation efficiency  & 0.404 \\ 
$\eta_\text{CHP}^\text{H}$         & CHP's thermal recovery efficiency  & 0.566 \\ 
$\eta_\text{B}$         & Boiler's thermal efficiency  & 0.900 \\ 
$\eta_\text{EV}^{\text{Ch}}$         & EV's charging efficiency  & 0.900 \\ 
$\eta_\text{EV}^{\text{DCh}}$         & EV's discharging efficiency  & 0.900 \\ 
$\eta_\text{TES}^{\text{Ch}}$         & TES's charging efficiency  & 0.900 \\ 
$\eta_\text{TES}^{\text{DCh}}$         & TES's discharging efficiency  & 0.900 \\ 
$S^\text{MAX}_{\text{TF}}$          & Transformer's Max. energy flow per hour     &   1000 kWh \\ 
$S^\text{MAX}_{\text{G}}$          & Natural Gas's Max. energy flow per hour     &   1200 kWh \\ 
$S^\text{MAX}_{\text{W}}$          & Wind Turbine's Max. energy flow per hour    &   200 kWh \\ 
$S^\text{MAX}_{\text{PV}}$          & Solar Panel's Max. energy flow per hour    &   200 kWh \\ 
$S^\text{MAX}_{\text{CHP}}$          & CHP's Max. energy flow per hour    &   300 kWh \\ 
$S^\text{MAX}_\text{B}$          & Boiler's Max.  energy flow per hour         &   800 kWh \\ 
$S^\text{Ch, MAX}_\text{EV}$          & EV's   Max. Charging energy flow per hour          &   80 kWh \\ 
$S^\text{DCh, MAX}_\text{EV}$          & EV's  Max. Discharging energy flow per hour         &   80 kWh\\ 
$S^\text{Ch, MAX}_\text{TES}$          & TES's Max. Charging energy flow per hour         &   50 kWh \\ 
$S^\text{DCh, MAX}_\text{TES}$          & TES's Max. Discharging energy flow per hour        &   50 kWh \\ 
$\text{SOC}^\text{MIN}_\text{EV}$          & EV's  Min. energy capacity         &   40 kWh \\ 
$\text{SOC}^\text{MAX}_\text{EV}$          & EV's  Max. energy capacity         &   80 kWh \\ 
$\text{SOC}^\text{MIN}_\text{TES}$          & TES's Min. energy capacity         &   40 kWh \\ 
$\text{SOC}^\text{MAX}_\text{TES}$          & TES's Max. energy capacity        &   200 kWh \\ \hline
\end{tabular}
\end{table}

\section{Case Studies}
\label{section cases}
To evaluate the proposed method, an IES with 4 EVs and 2 TESs is considered. The virtual replica of the IES is modelled as a MVES model that interacts with its physical twin, as well as the forecasting services and energy markets, as illustrated in Fig. \ref{fig data flow}. The time interval for each scheduling slot is 1 hour, thus one day has $T=24$ scheduling slots. {Note that this 1 hour time interval is selected as an example because the hourly forecasts and actual data are used in the case study, which can be adjusted according to specific system configurations.} Correspondingly, the input day-ahead forecasts $\mathbf{F}^t \triangleq \{{L}_\text{E}^{t}, {L}_\text{H}^{t}, S_\text{W}^{k,t}, S_\text{PV}^{k,t} \}$  are row vectors with a length of 96, while the output day-ahead scheduling $\mathbf{S}_\text{Sch}^t \triangleq \{S_\text{E}^{t}, S_\text{G}^{t}, S_\text{EV}^{k,t}, S_\text{TES}^{k,t}, \mathbf{v}^{t}\}$ are row vectors with a length of 216. A DNN with 3 hidden layers is used, where each hidden layer is implemented via a fully connected layer with the size of 768, 576 and 384, respectively. The Parametric ReLU function is exploited as the activation function for the input layer and each hidden layer, which returns $\max(0,x)+0.25 \min(0,x)$ for a given input $x$.
The outputs of the final fully connected layer are processed via the ``Constraint Enforcement'' module, which is illustrated in Fig. \ref{fig DNN structure} and specified in Section \ref{subsection Enforcing the Physical Constraints}. 

The day-ahead forecasts and real-time actual values of electricity loads, the wind energy generation and the solar energy generation are scaled from the U.K. historical records \cite{UK_E_Data_Lib}, while the heat loads are following the profile of domestic hot water consumption in \cite{gelavzanskas2015forecasting}, whose forecasting errors are generated with the profile in \cite{balint2019determinants}. 
The training dataset is generated as follows: a) for the forecast dataset, the full year's forecasts in 2019 are linearly combined in a random manner to form a dataset with a total of 56172 forecast vectors, and b) for the forecasting error dataset, the forecasts and actual values in 2019 are used to calculate the forecasting errors and a dataset with a length of 233 is generated, where the days with missing records and extreme errors (e.g., forecasting errors large than 45\%) are excluded.  The average day-ahead wholesale prices and real-time prices in the 2019 U.K. markets \cite{UK_E_Data_Lib}\cite{UK_Gas_Data_lib} are used, where $C_\text{E}^\text{DA}{=}\pounds 0.031/\text{kWh}$, $C_\text{G}^\text{DA}{=}\pounds 0.013/\text{kWh}$, $C_\text{E}^+{=}\pounds 0.058/\text{kWh}$, $C_\text{E}^-{=}\pounds 0.025{/}\text{kWh} $, $C_\text{G}^+{=}\pounds 0.022{/}\text{kWh}$ and $C_\text{G}^-{=}\pounds 0.011{/}\text{kWh}$. The rewarding rates are $C_\text{W} = \pounds 0.02{/}\text{kWh}$,  $C_\text{PV} = \pounds 0.02{/}\text{kWh}$, $C_\text{EV} = \pounds 0.03{/}\text{kWh}$ and $C_\text{TES} = \pounds 0.01{/}\text{kWh}$.

The proposed training method in Section III is used, which is illustrated in Fig. \ref{fig DNN structure}. The Adam optimiser \cite{adam_algorithm} with a learning rate of $10^{-5}$ is used to train the DNN in an unsupervised manner, where the penalty parameter $\lambda=1$ is used. To accelerate the training procedure, the mini-batch method is used, where the batch sizes for the day-ahead forecasts and the forecasting errors are empirically set as 4 and 55, respectively. During each training epoch, 10000 batches of day-ahead forecasts are used. {To evaluate the proposed method, the trained DNN is used for several case studies detailed in the following subsection, while the existing scheduling method based on day-ahead forecasts without forecasting error considerations is used as the benchmark algorithm.}

{To evaluate the proposed method, the trained DNN with the historical forecasts and forecast errors is used for the case studies to be detailed in the following subsections. Moreover, the day-ahead scheduling outputs of the DNN are used as the MVES scheduling decisions, while the real-time operation between the virtual replica MVES and the physical IES is detailed in Section \ref{subsection real-time operation}. }
  
{For the benchmark algorithm, the existing scheduling method based on day-ahead forecasts without forecasting error considerations is used. Specifically, the benchmark scheduling algorithm takes the day-ahead forecasts as inputs, and makes scheduling decisions so that for each scheduling slot, the energy supply by the IES meets the forecasted energy demands. 
}  

\subsection{A Case Study on the Hourly Cost Performance}
\begin{figure*}[!htbp]
\centering
\includegraphics[width=\FigSizeWide, trim={0cm 0.0cm 0cm 0.0cm},clip]{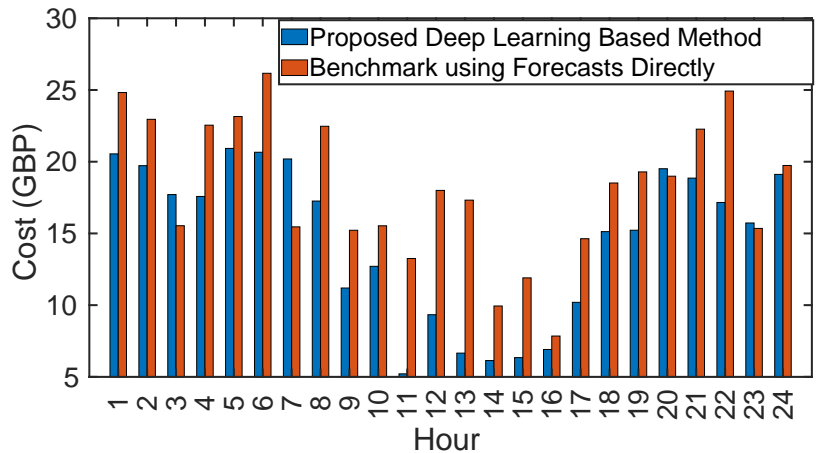}
\caption{{The hourly cost for 24 hours within one day, with 4 EVs and 2 TESs.} }
\label{fig Hourly}
\end{figure*}

The IES day-ahead scheduling outputs are the hourly scheduled operations for the CHPs, EVs and TESs, as well as the purchased electricity and natural gas from the day-ahead markets. Therefore the first case study is focused on the hourly cost performance, where the trained DNN via the proposed deep learning method is evaluated using a whole day's data (i.e., the complete 24 hours) in 2019, and the detailed hourly cost performance is reported in Fig. \ref{fig Hourly}. 

Comparing to the benchmark method, the proposed deep learning based method achieves better performance with reduced costs for most hours, while there are a few exceptions for the 3rd, 7th, 20th and 23rd hour, as illustrated in Fig. \ref{fig Hourly}.
This performance is expected, because the proposed method learns from historical errors and makes the statistically optimal scheduling decisions, i.e., the scheduling decision will: a) reduce the cost as much as possible for the most possible scenarios with the multiple uncertainty sources, and b) allow the cost to be increased for some scenarios while the long-term costs are still reduced. 
{Regarding the performance of executing time, the benchmark algorithm completes the day-ahead scheduling with an average of 0.355 s, while the proposed method completes the day-ahead scheduling with an average of 0.002 s. The millisecond-level execution time of the proposed method is because after the DNN is trained, the day-ahead scheduling decision can be obtained by feeding the day-ahead forecasts to the DNN. Since the NN parameters and structures are fixed, the execution time of inference is short.}
\subsection{A Case Study on the Daily Cost Performance}
\begin{figure*}[!htbp]
\centering
\includegraphics[width=\FigSizeWide, trim={0cm 0.0cm 0cm 0.0cm},clip]{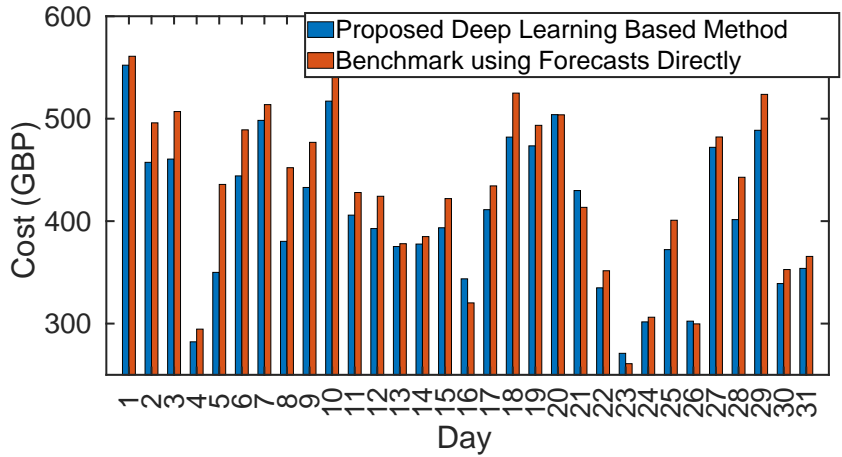}
\caption{{The daily cost for 31 days within one calendar month, with 4 EVs and 2 TESs.}}
\label{fig Daily}
\end{figure*}

The day-ahead scheduling decisions via the proposed method are for the 24-hour period (i.e., the whole day), where the IES devices are coordinated and scheduled to reduce the cost on a daily basis. Therefore in this case study, the daily cost performance is evaluated using the 31 days' data within one calendar month in May, 2019, whose results are presented in Fig. \ref{fig Daily}.  It can be seen that the proposed method is capable to reduce the daily costs for 27 days out of the total 31 days. This agrees well with the aim of the proposed method, which generally reduces the daily cost for the IES.

\subsection{A Case Study on the EV and TES Performance}
\label{subsection case study on EV and TES}
\begin{figure*}[!htbp]
\begin{minipage}{0.49\textwidth}
\centering
\includegraphics[width=\FigSize, trim={6cm 0.0cm 8cm 0.0cm},clip]{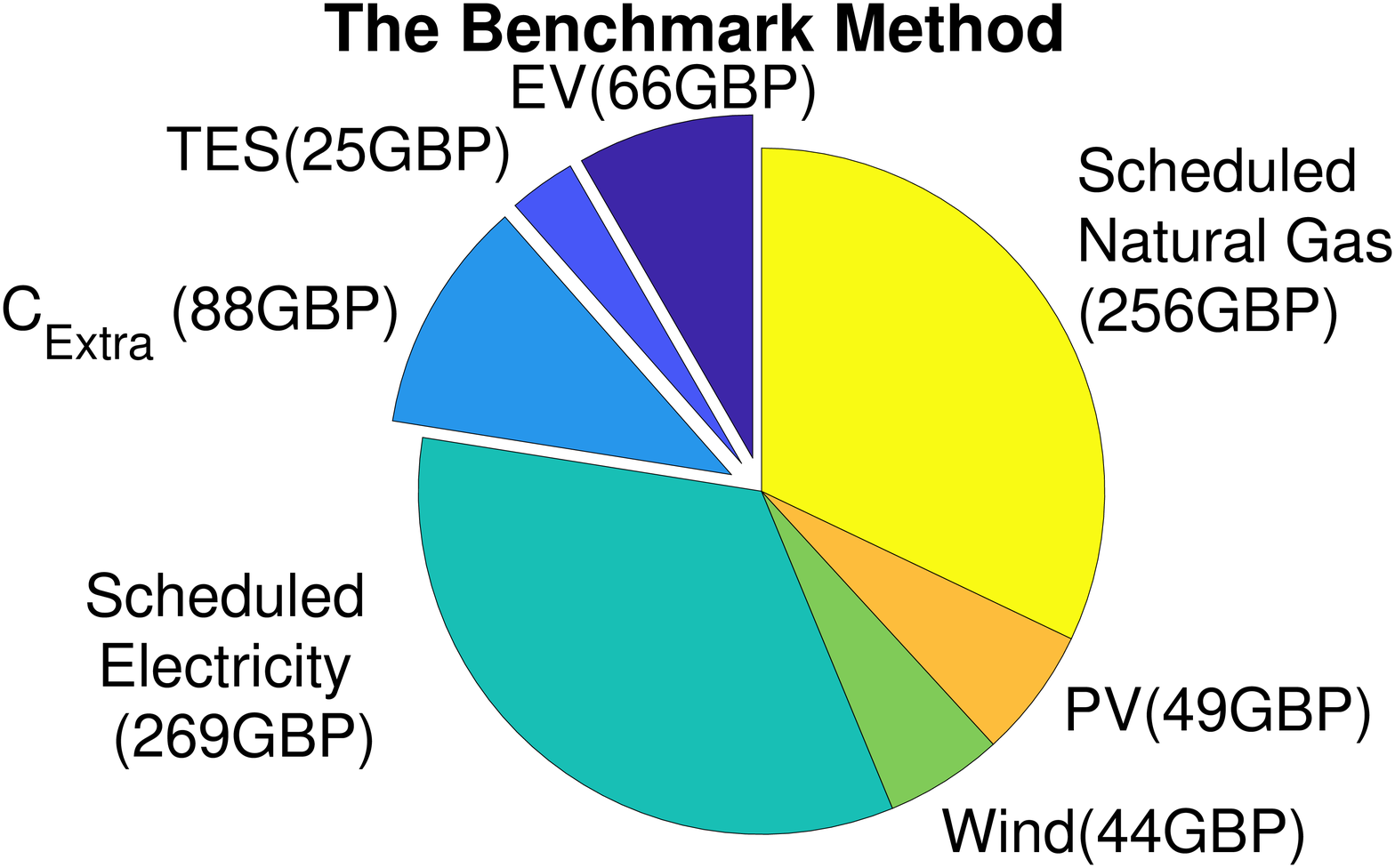}
\end{minipage}%
\begin{minipage}{0.49\textwidth}
\centering
\includegraphics[width=\FigSize, trim={6cm 0.0cm 7.5cm 0.0cm},clip]{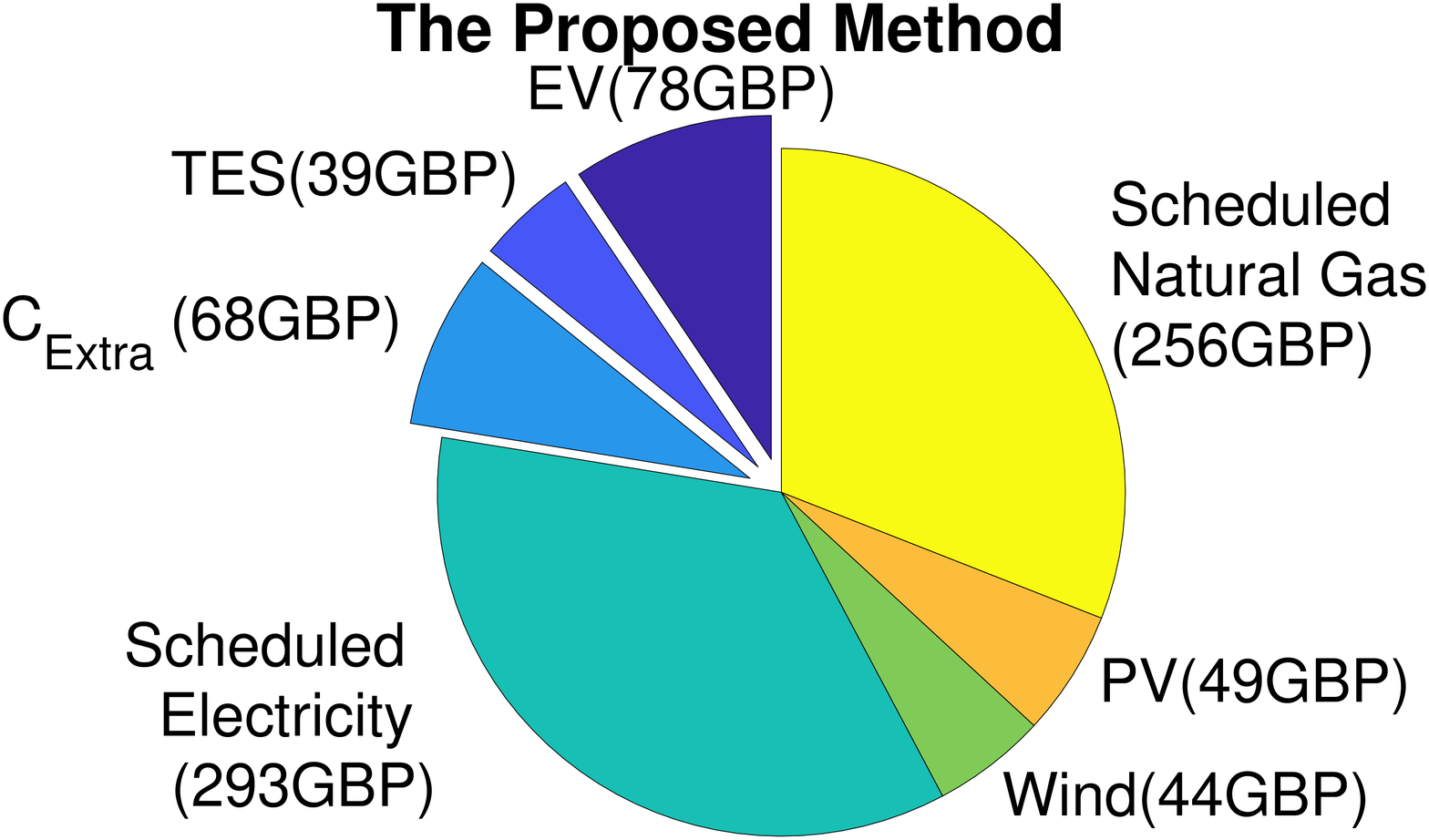}
\end{minipage}
\caption{The daily averaged cost performance of the benchmark method using forecasts directly and the proposed deep learning based method, with 4 EVs and 2 TESs.}
\label{fig ev tes pie bench}
\end{figure*}

The considered IES integrates multiple EVs and TESs, which are exploited as energy buffers in support of both electricity and thermal energy scheduling across the whole day. The averaged cost performance for each cost category is studied using 31 days' data in May, 2019, and the performance for the benchmark method and the proposed method are presented in Fig. \ref{fig ev tes pie bench}.

Comparing to the benchmark method, it can be seen from Fig. \ref{fig ev tes pie bench} that the proposed method pays more to the usage of EVs and TESs. Since EVs and TESs are paid by their absolute energy flows, it indicates the proposed method is scheduling more usage (charging or discharging) of the EVs and TESs across the day, comparing to the benchmark method. By jointly considering the reduced extra cost $C_\text{Extra}$ from Fig. \ref{fig ev tes pie bench}, as well as the daily cost reduction for most days from Fig. \ref{fig Daily}, it can be inferred that the proposed method shows a better performance in coordinating the EVs and TESs to help address the multiple uncertainty challenges, comparing to the benchmark method. The payments to the EVs and TESs are increased by 18.2\% and 56\%, respectively, which is beneficial to the IES as it will further encourage the participants of the EVs and TESs with a financial stimulation.

\subsection{A Case Study on the 2018 U.K. Dataset}
\begin{figure*}[!htbp]
\centering
\includegraphics[width=\FigSize, trim={0cm 0.0cm 0cm 0.0cm},clip]{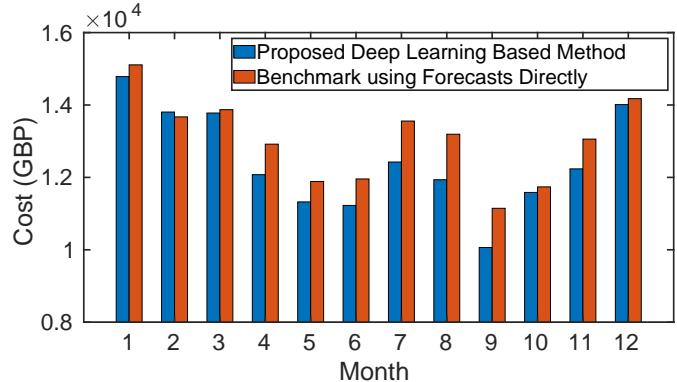}
\caption{{The monthly cost for the complete 12 months using 2018 U.K. dataset, with 4 EVs and 2 TESs.}}
\label{fig monthly}
\end{figure*}

During the training procedure of the proposed method, the training dataset for day-ahead forecasts and errors is generated from the U.K. historical data in 2019. In this case study, the trained DNN model will be evaluated using the U.K. historical data in 2018, in order to test if the obtained model is general to be applicable with the unknown dataset. The monthly cost for each calendar month in 2018 is reported in Fig. \ref{fig monthly}. It can be seen that the trained DNN model can still achieve the expected cost reduction performance, comparing to the benchmark method. It should be noticed that the proposed method exploits unsupervised training method, i.e., the DNN learns to reduce the cost without knowing the theoretical optimal day-ahead scheduling. This case study using U.K. data in 2018, together with the results in previous case studies, demonstrates that the proposed deep learning method is able to learn a practical and statistical solution for the day-ahead scheduling problem under multiple uncertain sources.    
 
The average daily costs in 2018 are detailed in Table \ref{table cost}, where the ideal case is calculated with no forecasting errors. Comparing to the ideal case, it can be calculated that the benchmark method pays a daily extra operating cost of \textsterling 34.5, while the proposed method pays a daily extra operating cost of \textsterling 12.6. \replaced{This means the proposed method reduces 63.5\% of the average daily extra operating cost induced by the multiple uncertainties, and a 5.1\% of the total average daily operating cost.}{This means the proposed method reduces the average daily extra operating cost by 63.5\%.}

\begin{table}[!htpb]
\centering
\caption{Daily Cost Performance with 4 EVs and 2 TESs}
\label{table cost}
\scriptsize
\begin{tabular}{ p{0.9cm} l l l l }
\hline
 & \textbf{\begin{tabular}[c]{@{}l@{}}Ideal \\ Case\end{tabular}} & \textbf{\begin{tabular}[c]{@{}l@{}}Proposed\\ Method\end{tabular}} & \textbf{\begin{tabular}[c]{@{}l@{}}Benchmark\\ Method\end{tabular}} & \textbf{\begin{tabular}[c]{@{}l@{}}Cost Due to \\ Physical Constraint\\  Adjustment\end{tabular}} \\ \hline
\textbf{\begin{tabular}[c]{@{}l@{}}Cost \\ (\textsterling)\end{tabular}} & 393.9 & 406.5 & 428.4 & 0.0012 \\ \hline
\end{tabular}
\end{table}

\subsection{A Case Study on Physical Constraint Adjustment Costs}
In this paper, the day-ahead scheduling decisions are outputs from a DNN, which should satisfy all the physical constraints as detailed in Section III. In the proposed method, the physical constraints regarding EVs' SOC and TESs' SOC are addressed via the loss function in \eqref{eq loss function g}, while the rest constraints are enforced via the ``Constraint Enforcement'' module embedded in DNN output layer. During the aforementioned case studies, the outputs of the trained DNN are further enforced to meet the EVs' SOC and TESs' SOC, where the scheduled EVs' energy flow and TESs' energy flow will be adjusted if their SOC are not met at any scheduled slot. This ensures the DNN satisfies all IES physical constraints for practical usage, but it could incur some additional costs since it is not as initially scheduled. As detailed in Table \ref{table cost}, it is seen the average additional cost due to physical constraint adjustment is \textsterling 0.0012, which is marginal to the daily costs. This also demonstrates that the proposed deep learning method is able to address the practical problems with multiple physical constraints.      

\subsection{A Case Study on Large System Setup}
{This} {case study evaluates the proposed method with a larger system setup, where the numbers of EVs and TESs are increased to 6 and 4, respectively. Besides, the thermal load demands have been increased by 20\%, which together with the increased number of EVs and TESs will test the effectiveness of the proposed method with the increased balance of supply and demand.}     

\begin{figure*}[!htbp]
\centering
\includegraphics[width=\FigSize, trim={0cm 0.0cm 0cm 0.0cm},clip]{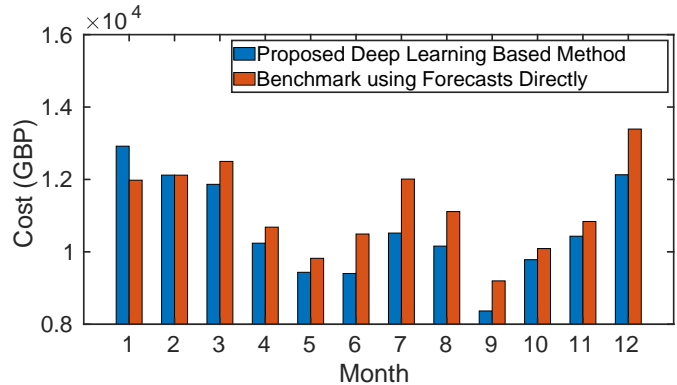}
\caption{{The monthly cost for the complete 12 months using 2018 U.K. dataset, with 6 EVs and 4 TESs.}}
\label{fig monthly E6T4}
\end{figure*} 

{We use the 2019 UK dataset as the training data and the 2018 UK dataset for testing. The experimental results are illustrated in Fig. \ref{fig monthly E6T4}. It can be observed that for 11 out of 12 months, the proposed method is able to reduce the monthly cost, comparing against the benchmark method. It also demonstrates the effectiveness of the proposed method under different system setups, e.g., more EVs and TESs as well as an increased balance of supply and demand. By jointly considering Fig. \ref{fig monthly}, it can be seen that for each month, the monthly cost has been much reduced in Fig. \ref{fig monthly E6T4}. This is because although the thermal load demands have been increased by 20\%, the effect of the operating cost reduction due to EVs and TESs is more prominent when their numbers are increased. Together with the case study on EVs and TESs in Section \ref{subsection case study on EV and TES}, the proposed method illustrates a better coordination strategy than the benchmark method.}

\section{Conclusions}
\label{section conclusions}
In this paper, a novel day-ahead scheduling scheme is proposed for the DT based IES. The proposed method enables active interactions between virtual replica and physical IES to form a DT system. Additionally, a deep learning method is proposed to make a statistically optimal day-ahead scheduling, by learning from historical forecasting errors and day-ahead forecasts. The challenging issue of multiple uncertainty sources is addressed via the proposed data augmentation based training method. Meanwhile, all physical constraints are enforced via both the designs of network architectures and the loss functions design. The performance is evaluated using case studies of historical U.K. data in 2018 and 2019. Comparing with the benchmark method using forecasts directly, the proposed method reduces the average daily extra operating cost by 63.5\%, which shows a promising solution to reduce the long-term IES operating cost. It can actively schedule EVs and TESs to better address the multiple uncertainty challenges in future energy systems, paving the way for realising net-zero target.

{For future research, we will further extend the proposed model and method to study their potential in carbon emission reduction. Moreover, the short-term forecasts and energy markets will be considered, where the proposed scheduling might be improved to further reduce the operation cost, e.g., by the intra-day rolling forecasts to update the scheduling decisions.
Besides, we will evaluate the proposed method in larger systems, including more equipment types, more EVs and TESs in numbers, and a larger balance of energy supply and demand.
\added{In addition, more advanced real-time scheduling to involve different IES facilities will be considered, which might lead to further reduced operating costs together with the day-ahead scheduling. } 
\added{Besides the random based data augmentation method, other general data augmentation methods will be investigated to better prepare the training data set and support the NN to learn the hidden forecasting error patterns.}

 \section*{Acknowledgement}
This work was supported by the Department for Business, Energy $\&$ Industrial Strategy (BEIS)  through the project of ``Ubiquitous Storage Empowering Response (USER)'' https://www.theuserproject.co.uk/.

\end{document}